\documentclass[12pt]{rspublic}

\usepackage[dvips,final]{graphicx}
\usepackage{epsfig}
\usepackage{epic}
\usepackage{eepic}
\usepackage{amssymb,amsmath}
\usepackage{times}
\usepackage{astron-jh}
\usepackage{subfigure}
\begin{document}

\newcommand{\vect}[1]{{ {\bf #1  }}} 
\newcommand{\uvect}[1]{{ \hat{\bf #1  }} }
\newcommand{\ci}{
	{
		{ {\bf c}}_i
	}
}
\newcommand{\fixme}[1]{

{ \bf{ ***FIXME: #1 }}

}
\newcommand{\half}{\frac{1}{2}}
\newcommand{\cop}{\Omega_i^\sigma}
\newcommand{\copbgk}{{\Omega_i^\sigma}_{\mathrm{BGK}}}
\newcommand{\Dop}{\mathbb D}
\newcommand{\tausig}{\tau^\sigma}
\newcommand{\xsig}{x^\sigma}
\newcommand{\tausigb}{\tau^\bar{\sigma}}
\newcommand{\psisig}{\psi^\sigma}
\newcommand{\psisigb}{\psi^{\bar{\sigma}}}
\newcommand{\nusig}{\nu^\sigma}
\newcommand{\msig}{m^\sigma}
\newcommand{\nsig}{n^\sigma}
\newcommand{\usig}{u^\sigma}
\newcommand{\usiga}{u^\sigma_\alpha}
\newcommand{\Fsig}{F^\sigma}
\newcommand{\Fsiga}{F^\sigma_\alpha}
\newcommand{\upr}{u'}
\newcommand{\upra}{{u'}_\alpha}
\newcommand{\vsig}{v^\sigma}
\newcommand{\vsiga}{v^\sigma_\alpha}
\newcommand{\sumsig}{\sum_\sigma}
\newcommand{\sumsigb}{\sum_{\bar{\sigma}}}
\newcommand{\sumsigsigb}{\sum_{\sigma\bar{\sigma}}}
\newcommand{\sumi}{\sum_i}
\newcommand{\msi}{\msig\sum_i}
\newcommand{\ciao}{c_{i{\alpha_1}}}
\newcommand{\cian}{c_{i{\alpha_n}}}
\newcommand{\cia}{c_{i\alpha}}
\newcommand{\cib}{c_{i\beta}}
\newcommand{\cig}{c_{i\gamma}}
\newcommand{\cid}{c_{i\delta}}
\newcommand{\cs}{c_{\mathrm{s}}}
\newcommand{\rhosig}{\rho^\sigma}
\newcommand{\frt}{\frac{\rhosig}{\tausig}}
\newcommand{\xt}{(\vect{x},t)}
\newcommand{\xpct}{(\vect{x}+\ci,t)}
\newcommand{\xpc}{(\vect{x}+\ci)}
\newcommand{\Ua}{U_\alpha}
\newcommand{\fis}{f_i^\sigma}
\newcommand{\fisb}{\bar{f}_i^\sigma}
\newcommand{\Nis}{N_i^\sigma}
\newcommand{\NiU}{N_i^\sigma({\bf U})}
\newcommand{\Niu}{N_i^\sigma({\bf u})}
\newcommand{\Nivs}{N_i^\sigma({\bf v}^\sigma)}
\newcommand{\Ti}{T_i}
\newcommand{\Ta}{{T}^{(1)}_{\alpha}}
\newcommand{\Tab}{{T}^{(2)}_{\alpha\beta}}
\newcommand{\Tabg}{{T}^{(3)}_{\alpha\beta\gamma}}
\newcommand{\kronab}{\delta_{\alpha\beta}}
\newcommand{\kronag}{\delta_{\alpha\gamma}}
\newcommand{\kronbg}{\delta_{\beta\gamma}}
\newcommand{\oz}[1]{{#1}^{(0)}}
\newcommand{\oo}[1]{{#1}^{(1)}}
\newcommand{\ot}[1]{{#1}^{(2)}}
\newcommand{\ordn}[1]{{#1}^{(n)}}
\newcommand{\partiald}[2]{
	\frac { \partial #1 } { \partial #2 }
}
\newcommand{\partialdd}[2]{
	\frac { \partial^2 #1 } { \partial {#2}^2 }
}

\newcommand{\partialop}[1]{
	\frac { \partial } { \partial #1 }
}
\newcommand{\partialopop}[1]{
	\frac { \partial^2 } { \partial {#1}^2 }
}
\newcommand{\dal}{\partial_\alpha}
\newcommand{\dbe}{\partial_\beta}
\newcommand{\dga}{\partial_\gamma}
\newcommand{\dt}{\partial_{t}}
\newcommand{\dit}{\partial_{1t}}
\newcommand{\dtt}{\partial_{2t}}
\newcommand{\Psa}{\Pi^\sigma_\alpha}
\newcommand{\Psab}{\Pi^\sigma_{\alpha\beta}}
\newcommand{\Psabg}{\Pi^\sigma_{\alpha\beta\gamma}}
\newcommand{\Pa}{\Pi_\alpha}
\newcommand{\Pab}{\Pi_{\alpha\beta}}
\newcommand{\Pabg}{\Pi_{\alpha\beta\gamma}}
\newcommand{\ep}{\epsilon}

\title{Large-scale grid-enabled lattice-Boltzmann simulations of complex
fluid flow in porous media and under shear}

\author[J.~Harting, M.~Venturoli and P.V.~Coveney]{Jens Harting$^1$, Maddalena Venturoli$^{1,2}$ and Peter V. Coveney$^1$}

\affiliation{$^1$Centre for Computational Science, Christopher Ingold Laboratories, University College London, 20 Gordon Street, London WC1H 0AJ, UK
\\ $^2$Schlumberger Cambridge Research, High Cross, Madingley Road, Cambridge CB3 0EL, UK}

\label{firstpage}

\maketitle

\begin{center}
\begin{abstract}{Lattice-Boltzmann, porous media, complex fluids under shear,\\ grid
computing, computational steering}
Well designed lattice-Boltzmann codes exploit the essentially
embarrassingly parallel features of the algorithm and so can be run with
considerable efficiency on modern supercomputers. Such scalable codes
permit us to simulate the behaviour of increasingly large quantities of
complex condensed matter systems.  In the present paper, we present some
preliminary results on the large scale three-dimensional lattice-Boltzmann
simulation of binary immiscible fluid flows through a porous medium
derived from digitised x-ray microtomographic data of Bentheimer
sandstone, and from the study of the same fluids under shear.  Simulations
on such scales can benefit considerably from the use of computational
steering and we describe our implementation of steering within the
lattice-Boltzmann code, called LB3D, making use of the RealityGrid
steering library. Our large scale simulations benefit from the new concept
of capability computing, designed to prioritise the execution of big jobs
on major supercomputing resources.  The advent of persistent computational
grids promises to provide an optimal environment in which to deploy these
mesoscale simulation methods, which can exploit the distributed nature of
compute, visualisation and storage resources to reach scientific results
rapidly; we discuss our work on the grid-enablement of
lattice-Boltzmann methods in this context.
\end{abstract}
\end{center}

\section{Introduction}
The length and time scales that can be modelled using microscopic
modelling techniques such as molecular dynamics are circumscribed by the
limited computational resources available today. Even with today's fastest
computers, the accessible length scales are on the order of nanometres and
the time scales restricted to the nanosecond range. Mesoscopic models open
the way to studies of time dependent, non-equilibrium phenomena occurring
in much larger systems and on orders of magnitude longer timescales, thus
bridging scales between microscopic models and macroscopic or continuum
approaches. 

In this paper, we use the lattice-Boltzmann method to model binary fluids
under shear and flow in a porous medium. In the porous medium case, we are
now able to reach length scales for the simulated fluid flow which can be
compared directly to data gleaned from magnetic resonance imaging
experiments. In the case of fluids under shear, one must take care of
finite size effects which make it undesirable to study cubic systems, but
rather preferable to study systems of high aspect ratio. In this paper we 
present preliminary results, but expect to return with extensive descriptions
of the two applications in the near future.  Both problems are
very computationally demanding and require today's top-of-the-range
supercomputers and large scale data storage facilities. Since these
resources are expensive, we have to handle them with care and minimize
wastage of CPU time and disk space. A good initial position is to make
sure that our simulations do not last longer than needed and do not
produce more data than necessary. But more than this, computational
steering allows us to interact with a running simulation and adjust
simulation parameters and data dumping rates; it also enables us to
monitor the state of our simulations and react immediately if they do not
behave as expected, as we shall discuss later. 
The world wide effort to develop reliable computational grids gives us
hope to run our simulations in an even more efficient way.  Computational
grids are a collection of geographically distributed and dynamically
varying resources, each providing services such as compute cycles,
visualization, storage, or even experimental facilities.  It is hoped that
computational grids will offer for information technology what electricity
grids offer for other aspects of our daily life: a transparent and reliable resource that
is easy to use and conforms to commonly agreed standards \cite{gridbook2,bib:Berman}. 
Then we shall be able to use the available compute 
resources in a transparent way,
leaving to smart middleware the task of finding the best available machines
to run simulations on and migrate them to other platforms if
necessary to ensure optimal performance. Grids will also allow 
storage, compute and visualization resources to be widely distributed
without our having to care about their location.

The main purpose of the present paper is to introduce the concepts of
computational steering and grid computing to an audience of computational
scientists, concerned here with simulation of fluid dynamics. The paper is
structured as follows.  After a short introduction to our lattice-Boltzmann 
method in Section \ref{Sec:LatticeBoltzmann}, we give a
description of our implementation of computational steering in Section
\ref{Sec:CompSteer} and explain the advantages we expect to gain from the
advent of computational grids in Section \ref{Sec:Grids}. Sections
\ref{Sec:Shear} and \ref{Sec:Porous} contain our preliminary results on
large scale grid-enabled simulations of fluids under shear and in porous
media. We present our conclusions in Section \ref{Sec:Conclusion}.

\section{A lattice-Boltzmann model of immiscible fluids}
\label{Sec:LatticeBoltzmann}
\newcommand{\fone}{f_1(\vect{r},\vect{v},t)}
During the last decade, many authors have shown that the lattice-Boltzmann
algorithm is a powerful method for simulating fluid dynamics. 
This success is due to its
simplicity and to facile computational implementations \cite{bib:chin-harting-jha,bib:love-nekovee-coveney-chin-gonzalez-martin,bib:nekovee-chin-gonzalez-coveney,bib:succi}. Instead
of tracking individual atoms or molecules, the lattice-Boltzmann method
describes the dynamics of the single-particle distribution function of
mesoscopic fluid packets.

In a continuum description, the single-particle distribution function
$\fone$ represents the density of fluid particles with position
$\vect{r}$ and velocity $\vect{v}$ at time $t$, such that the density
and velocity of the macroscopically observable fluid are given by
$\rho(\vect{r},t) = \int \fone {\mathrm d}\vect{v} $ and
$\vect{u}(\vect{r},t) = \int \fone \vect{v} {\mathrm d} \vect{v}$
respectively. In the non-interacting, long mean free path limit, with no
externally applied forces, the evolution of this function is described
by the Boltzmann equation

\begin{equation}
\label{eq:boltzmann}
\left( \dt + \vect{v} \cdot \vect{\nabla} \right) f_1
= \Omega[f_1].
\end{equation}
While the left hand side describes changes in the distribution function
due to free particle motion, the right hand side models pairwise
collisions. This collision operator $\Omega$ is an integral expression
that is often simplified~\cite{bib:bgk} to the linear
Bhatnagar-Gross-Krook (BGK) form
\begin{equation}
\label{eq:bgk}
\Omega[f] \simeq - \frac 1 \tau \left[ f - f^{\mathrm{(eq)}} \right].
\end{equation}
This collision operator describes the relaxation, at a rate controlled by
a characteristic time $\tau$, towards a local Maxwell-Boltzmann equilibrium
distribution $f^{\mathrm{(eq)}}$. It can be shown that distributions
governed by the simple Boltzmann-BGK equation conserve mass, momentum, and
energy \cite{bib:succi}. They obey a non-equilibrium form of the second law of
thermodynamics~\cite{bib:liboff} and the Navier-Stokes equations for
macroscopic fluid flow are obeyed on coarse length and time scales
~\cite{bib:chapman-cowling,bib:liboff}.

By discretizing the single-particle distribution function in space and time, one
obtains the usual lattice-Boltzmann formulation, where the positions $\vect{r}$
on which $\fone$ is defined are restricted to points $\vect{r}_i$ on a
Bravais lattice. The velocities $\vect{v}$ are restricted to a set $\ci$ joining
points on the lattice and the density of particles at lattice site
$\vect{r}$ travelling with velocity $\ci$, at timestep $t$ is given by
$f_i(\vect{r},t) = f(\vect{r},\ci,t)$.  The fluid's density and velocity
are given by

\begin{equation}
\label{eq:lbe-density}
\rho(\vect{r}) = \sum_i f_i(\vect{r})
\end{equation}

\begin{equation}
\label{eq:lbe-velocity}
\vect{u}(\vect{r}) = \sum_i f_i(\vect{r}) \ci
\end{equation}

The discretized Boltzmann description can be evolved as a two-step
procedure. In the collision step, particles at each lattice site are
redistributed across the velocity vectors; this process corresponds to the
action of the collision operator.
In the advection step, values of the post-collisional distribution
function are propagated to adjacent lattice sites.

By combining the two steps, one obtains the lattice-Boltzmann equation
(LBE)

\begin{equation}
\label{eq:lbgk}
f_i(\vect{r},t+1) - f_i(\vect{r},t)
= \Omega[f] \\
= - \frac 1 \tau \left[ f_i(\vect{r},t) 
- N_i\left( \rho, \vect{u} \right)\right],
\end{equation}
where $N_i = N_i\left(\rho(\vect{r}),\vect{u}(\vect{r})\right)$ is a
polynomial function of the local equilibrium density and velocity, and can be found by
discretizing the Maxwell-Boltzmann equilibrium distribution.

Our lattice-Boltzmann implementation uses the Shan-Chen
approach~\cite{bib:shan-chen}, by incorporating an explicit forcing term
in the collision operator in order to model multicomponent interacting
fluids. Shan and Chen extended the single-particle distribution function
$f_i$ to the form $f_i^\sigma$, where each component is denoted by a
different value $\sigma$, so that the density and momentum of a single
component $\sigma$ are given by $\rhosig = \sumi \fis$ and $\rhosig
\vect{u}^\sigma = \sumi \fis \ci$ respectively. The fluid viscosity
$\nu^\sigma$ is proportional to $(\tau^\sigma-1/2)$ and the mass of each
particle is $m^\sigma$. This results in a
lattice BGK equation (\ref{eq:lbgk}) of the form
\begin{equation}
\label{eq:lbgk-sc}
\fis(\vect{r},t+1) - \fis(\vect{r},t) =
- \frac 1 \tausig
\left[
\fis - N_i(\rhosig, \vect{v}^\sigma)
\right]
\end{equation}
The velocity $\vect{v}^\sigma$ is found by calculating a weighted average
velocity 
\begin{equation}
\vect{u}' = \left( \sumsig \frac \rhosig \tausig \vect{u}^\sigma
\right)
/ \left( \sumsig \frac \rhosig \tausig \right),\end{equation}
and then adding a term $\vect{F}^\sigma$ to account for additional forces,
\begin{equation}
\vect{v}^\sigma = \vect{u}' + \frac \tausig \rhosig \vect{F}^\sigma .
\end{equation}
To produce nearest-neighbour interactions between fluid
components, this term assumes the form
\begin{equation}
\label{eq:colour-colour}
\vect{F}^\sigma = 
- \psisig ( \vect{x} )
\sumsigb g_{\sigma \bar{\sigma}}
        \sumi \psisigb \left( \vect{x} + \ci \right) \ci,
\end{equation}
where $\psisig ( \vect{x} ) = \psisig ( \rhosig ( \vect{x}))$
is an effective charge for component $\sigma$, set equal to the fluid
component density, that is $\psisig ( \vect{x} )=\rhosig( \vect{x} )$ ; $g_{\sigma \bar{\sigma}}$
is a coupling constant controlling the strength of the interaction
between two components $\sigma$ and $\bar{\sigma}$. If $g_{\sigma
\bar{\sigma}}$ is set to zero for $\sigma = \bar{\sigma}$, and to a
positive value for $\sigma \neq \bar{\sigma}$ then, in the interfacial
region between bulk domains of each component, particles experience a force in
the direction away from the interface, producing immiscibility. For
two-component systems, we use the notation $g_{\sigma
\bar{\sigma}} = g_{\bar{\sigma}\sigma} = g_{br}$.
External forces are added in a similar manner. For example, in order to produce a gravitational force acting in the $z$-direction, the
force term $\vect{F}^\sigma$ can take the form $g \rho^\sigma
\hat{\vect{z}}$. 

A convenient way to characterize binary fluid mixtures is in terms of the order
parameter or colour field
\begin{equation}
\phi(\vect{x}) =
\rho^r(\vect{x})-\rho^b(\vect{x}).
\label{Eq:Colour}
\end{equation}
The order parameter is positive in areas of high concentration of `red'
fluid and negative in areas of `blue' dominance; the
isosurface $\phi(\vect{x}) = 0$ denotes the interface between both fluid
constituents.

The model has been extended to handle amphiphiles, which are treated as
massive point like dipoles with different interaction strengths on each
end and a rotational degree of freedom~\cite{bib:chen-boghosian-coveney}.
Our code, LB3D, can handle binary and ternary fluid mixtures with or
without amphiphiles. But since we only discuss simulations of binary
fluids in this paper, we refer the reader to other papers
\cite{bib:chen-boghosian-coveney,bib:nekovee-chin-gonzalez-coveney,bib:love-nekovee-coveney-chin-gonzalez-martin,bib:chen-boghosian-coveney} for a
more comprehensive description of the amphiphilic case. 

\section{Computational steering of lattice-Boltzmann simulations}
\label{Sec:CompSteer}
This section outlines the benefits of computational steering for high
performance computing applications. Our three-dimensional lattice-Boltzmann
code (LB3D) and its computational steering implementation are used to
illustrate the substantial improvements which computational steering offers in
terms of resource efficiency and time to discover new physics.

Traditionally, large, compute-intensive simulations are run non-interactively.
A text file describing the initial conditions and parameters for the course of
a simulation is prepared, and then the simulation is submitted to a batch queue
on a large compute resource. The simulation runs entirely according to the
prepared input file, and outputs the results to disk for the user to copy
to his local machine and examine later.

This mode of working is sufficient for many simple investigations of mesoscale
fluid behaviour, but has several drawbacks. Firstly, consider the situation
where one wishes to examine the dynamics of the separation of two immiscible
fluids: this is a subject which has been of considerable interest in the
modelling community in recent
years~\cite{bib:gonzalez-nekovee-coveney,bib:kendon-cates-pagonabarraga-desplat-bladon}.
Typically, a guess is made as to how long the simulation must run before
producing a phase separation, and then the code is run for a fixed number of
timesteps. If a phase transition does not occur within this number of
timesteps, then the job must be resubmitted to the batch queue, and restarted.
However, if a phase transition occurs in the early stages of the simulation,
then the rest of the compute time will be spent simulating an equilibrium
system of very little interest. Even worse, the initial parameters of the
system might turn out not to produce a phase separation at all and all of the
CPU time invested in the simulation will have been wasted.

Computational steering is a way to overcome these drawbacks. It allows the
scientist to interact with a running simulation and to change or monitor
simulation parameters on the fly. Examples of monitored parameters are the
timestep, surface tension, density distributions, or `colour' fields.
Steerable parameters are data dumping frequencies, relaxation times or
shear rates. One can also `stop', `pause' or `restart' from a previously
saved checkpoint. A `checkpoint' is a set of files representing the state
of the simulation and allowing the code to be restarted without rerunning
earlier steps of the simulation.  The `restart' functionality is
particularly important since it provides the basis of a system that allows
the scientist to `rewind' a simulation. Having done so, it can then be run
again, perhaps after having steered some parameter or altered the
frequency with which data from the simulation is recorded. 

We have implemented computational steering within the LB3D code with the help of
colleagues at Manchester Computing as part of the ongoing RealityGrid
project (http://www.realitygrid.org) \cite{bib:chin-harting-jha,bib:brooke-coveney-harting,bib:coveney-nobel}.
The RealityGrid project aims to enable the modelling and simulation of
complex condensed matter structures at the molecular and mesoscale levels
as well as the discovery of new materials using computational grids. The
project also involves biomolecular applications and its long term ambition
is to provide generic computational grid based technology for scientific,
medical and commercial activities.

Within RealityGrid, computational steering has been implemented in order
to enable existing scientific computer programs (often written in
Fortran90 and designed for multi-processor/parallel supercomputers) to be
made steerable while minimising the amount of work required. The steering
software has been implemented as a library written in C and is thus
callable from a variety of languages (including C, C++ and Fortran90). The
library completely insulates the application from any implementation
details. For instance, the process by which messages are transferred
between the steering client and the application (e.g. via files or
sockets) is completely hidden from the application code and the steering
library does not assume or prescribe any particular parallel-programming
paradigm (such as message passing or shared memory). The steering protocol
has been designed so that the use of steering is never critical to the
simulation. Thus, a steering client can attach and detach from a running
application without affecting its state.

The ability to monitor the state of a simulation and use this to make steering
decisions is very important. While a steering client provides some information
via the simulation's monitored parameters, a visualisation of some aspect of
the simulation's state is often required. In our case this is usually a
three-dimensional data set, visualised by a second software component using
isosurfacing or volume rendering. 

The steering library itself consists of two parts: an application side and a
client side. Using a generic steering client written in C++ and Qt (a
GUI toolkit, http://www.trolltech.org) one is capable to steer any application
that has been `steering enabled' using the library.

The RealityGrid steering library and client are generic enough to be
interfaced to by almost any simulation code. Usually only a couple of
hours have to be invested in adapting a code to do simple parameter
steering and monitoring. Indeed, since the initial version of the steering
library was written at least four other codes have been made steerable in
this way (these include molecular dynamics, Monte Carlo and other lattice-Boltzmann
codes).

In addition to the features the steering library provides, LB3D has its own
logging and replay facilities which permit the user to `replay' a steered
simulation. This is an important feature since it allows the data from
steered simulations to be reproduced without human intervention. Moreover,
this feature can be used as an `auto-steerer'. Thus multiple simulations,
which read different input files at startup and are `steered' in the same
way, can be launched without the need for human intervention during the
simulation. One application of this particular feature appears in 
studies of how changes in parameters affect a simulation that has evolved for
a given number of timesteps. Another application is the automatic adaptation
of data dumping or checkpointing frequencies. If the user has found from a
manually steered simulation that no effects of interest are expected for a
given number of initial timesteps, he can reduce the amount of data
written to disk for early times of the simulation.

A more detailed description of computational steering and its
implementation within RealityGrid can be found in recently published
papers \cite{bib:chin-harting-jha,bib:brooke-coveney-harting}. 
\cite{bib:chin-harting-jha}
also contains an example demonstrating the usefulness of
computational steering of three dimensional lattice-Boltzmann simulations:
parameter searches are a common task we have to handle because our 
lattice-Boltzmann method has a range of free parameters - only by choosing them
correctly, can one simulate effects of physical interest. Previously,
these parameter searches have been performed using a taskfarming approach:
many small simulations with different parameters have been launched. In
such cases we have used up many thousands of CPU hours and needed hundreds
of gigabytes of simulation data to be stored for a single large scale
parameter search. Computational steering offers the possibility to
`joystick' through parameter space in order to find regions of interest.
In this way, the resources needed can be substantially reduced. The main
benefit, however, is the reduced amount of data that has to be analysed
subsequently
since this is the most time consuming and demanding task. While
simulations can be completed within days or weeks, analysis usually takes
months. We have also found that if the amount of data to be searched for
interesting effects exceeds certain bounds, it is almost impossible for a
human to keep track of it. One might suggest automation of the analysis
process, but the human eye turned out to be the only reliable tool for our
simulations. It is often very easy to spot effects occurring in data sets
by looking at isosurfaces or volume-rendered visualisations; by contrast
automation of the analysis of the generated data is much harder because it
can be difficult to define the effects sufficiently well, impossible
to anticipate the effects sought in advance, or simply not worthwhile to
invest additional effort in the development of algorithms to automate the
process. 

\section{Capability computing and terascale computational grids}
\label{Sec:Grids}
Three-dimensional lattice-Boltzmann simulations are very computationally
demanding and need high performance computing resources (i.e.
supercomputers). In order to reach length and time scales which can be
compared with experimental data and to eliminate finite size effects, one
needs large lattices, for example 512$^3$ or 1024$^3$.  Simulations also
have to run for several thousands or tens of thousands of time steps, thus
pushing the required compute and storage resources beyond what is
typically available to users on medium scale supercomputers today.  In the
case of LB3D, the main restriction is the per CPU memory available, which
on all machines we have access to currently is not more than 1GB. For
example, we require at least 1024 CPUs to simulate a 1024$^3$ system. 

Obviously, computational steering becomes an even more useful tool here
because large scale simulations are very expensive; it is essential that
the simulation does not generate useless data, and that the expensive
resources are used as efficiently as possible. 
The need for vast compute power has brought with it the concept of
`capability computing'. We understand this term as a description of how
large jobs are handled by supercomputing centres: large jobs are favoured
and assigned a higher priority by the queueing system. In these terms,
`large' refers to jobs that request at least half of the total number of
CPUs available. With standard job queue configurations operating on batch
systems, there is a strong disincentive to submit large jobs: if a user
submits a `large' job, turn around times can be very long, making such
high end resources incompetitive compared to modern commodity clusters
which are becoming widely available locally.  In some cases,
supercomputing centres can offer discounts for large (capability
computing) jobs if the simulation code  can be shown to scale well. LB3D
has recently been awarded a gold-star rating for its excellent scaling
capabilities by the HPCx Consortium (http://www.hpcx.ac.uk) allowing us to
run simulations on 1024 CPUs (the full production partition) of their 1280
CPU IBM SP4, with a discount of 30\% \cite{bib:harting-wan-coveney}. The
flow in porous media simulations described later in this chapter have been
done on up to 504 CPUs of a 512 CPU SGI Origin 3800 at the CSAR service in
Manchester, UK (http://www.csar.cfs.ac.uk). LB3D scales linearly on all
platforms available to us.  In addition to those mentioned above, these
include a CRAY T3E, a 3000 CPU Compaq Alpha cluster `Lemieux' (at
Pittsburgh Supercomputing Center), various Linux clusters and SGI Origin
2000 and 3800 systems. 

We expect it to become easier to simulate systems of a size which is
comparable to experimental data with the advent of computational grids
\cite{gridbook2,bib:Berman}.  Grids are geographically distributed and
dynamically varying collections of resources such as supercomputers,
storage facilities or advanced experimental instruments that are connected
by high speed networks, thus allowing widespread human collaborators to
work together closely. In the same way that large scale lattice-Boltzmann
simulations require a supercomputer, the visualisation of the large and
complex data sets that these simulations produce also require specialist
hardware that few scientists have direct access to. As we shall see,
visualisation engines can also be treated as distributed resources on the
grid.

Computational grids are related to traditional distributed computing, with
the major extension that they enable the transparent sharing and
collective use of resources, which would otherwise be individual and
isolated facilities.  With growing intensity, significant effort is being
invested worldwide in grid computing (http://www.gridforum.org).  The grid
aims to present the elements required for a computational task (e.g.
calculation engine,  filters, visualisation capability) as components
which can be effectively and transparently coupled through the grid
framework using middleware. In this scenario, any application or simulation
code can be viewed simply as a data producing or consuming object on the
grid and computational steering is a way of allowing users to interact
with such objects.

Our lattice-Boltzmann code LB3D is now a fully grid-enabled, steerable
application. LB3D simulations can then be launched and steered on a remote
machine, with the visualisations being performed in other geographic
locations. One or more users control the workflow from a laptop running a
steering client and client software to interact remotely with the compute
and visualisation engines. Behind the scenes, the `grid middleware' moves
files, simulation data and commands between the resources involved. We
have used both  Globus~\cite{globus} (http://www.globus.org) and
Unicore~(http://www.unicore.org) as the basic middleware fabric in this
work, and digital certificates provided by the UK e-science certification
authority (http://www.grid-support.ac.uk). 

The grids being used in these demonstration activities have been assembled
especially for each event. By contrast, the  UK e-Science  community has
constructed an ambitious Level 2 Grid ~(http://www.grid-support.ac.uk/l2g)
that aims to provide the user community with a persistent grid of
heterogenous resources. LB3D and the RealityGrid steering framework have
already been deployed on this Level 2 Grid, which uses Globus GT2 as
middleware. Thus we are amongst the first groups in the world to use a
persistent grid for scientific research requiring high performance
computing and computational steering.

In a major US/UK grid project leading up to and including Supercomputing
2003 we are studying the defect formation and dynamics within a
self-assembled gyroid mesophase
\cite{bib:gonzalez-coveney} utilizing a fast network between the `Extended
Terascale Facility'~(http://www.teragrid.org) in the USA and the national
supercomputing centres at Manchester (CSAR) and Daresbury (HPCx) in the
UK. This gives us access to machines in the US including Lemieux and
various Itanium (IA64) systems. In the UK, access to the total combined
resources of CSAR and HPCx provides us with a 1280 CPU IBM SP4
machine, a 504 CPU SGI Origin 3800, and a 256 CPU SGI Altix. For
visualization there are resources available on both sides of the Atlantic
as well, including various SGI Onyx machines and commodity clusters which
use `Chromium' (http://chromium.sourceforge.net) for parallel rendering.
The Visualisation Toolkit (VTK, http://www.kitware.com) allows us to
generate isosurfaces or volume-rendered visualisations of even our largest
data sets on these platforms (see http://www.realitygrid.org/TeraGyroid.html).

However, the vision of a computational grid that furnishes for information
technology what electricity (and other utility) grids have achieved in
terms of almost universal and transparent access to energy (and other
resources) within modern civilised societies remains a dream today. Many
problems remain to be addressed before computational grids become easy to
use. At the time of writing, it remains very awkward both to access and
utilise grid-enabled resources and the much vaunted advantages are yet to
be realised.  In fact, our own experiences indicate that real progress
towards usablity will be achieved most quickly with the development and
deployment of leightweight middleware, in marked contrast with the
existing heavyweight behemoths.

Moreover, for effective utilisation of computational steering together
with large scale simulations, it is very important that supercomputing
centres change their policy of job scheduling since advanced reservation
for the co-allocation of compute and visualisation resources becomes
essential.  Today, this is possible for small scale simulations which do
not run on the grid if turn around times are short, but for large scale
jobs one needs special arrangements with the resource owners. It is also
important that users will be able to request access to resources at
convenient times, i.e. during working hours rather than in the middle of
the night. We expect that the huge effort currently being invested in the
development of grid standards will result in a satisfactory solution of
these issues.  We believe that computational grids will revolutionise the
way scientific simulations are performed in the near future because they
should then offer an easy and effective way to access distributed
resources in an optimal way for the scientific problem under
investigation. 

\section{Immiscible fluid mixtures under shear}
\label{Sec:Shear}
Lees and Edwards published their method for the application of shearing in
molecular dynamics simulations in 1972 \cite{bib:lees-edwards}. Since then
Lees-Edwards boundary conditions have become a popular method for simulating fluid
rheology under shear using a variety of different methods and have
been implemented in lattice-Boltzmann codes before
\cite{bib:wagner-yeomans-shear,bib:wagner-pagonabarraga}. 

\begin{figure}
\centerline{\includegraphics[width=15cm]{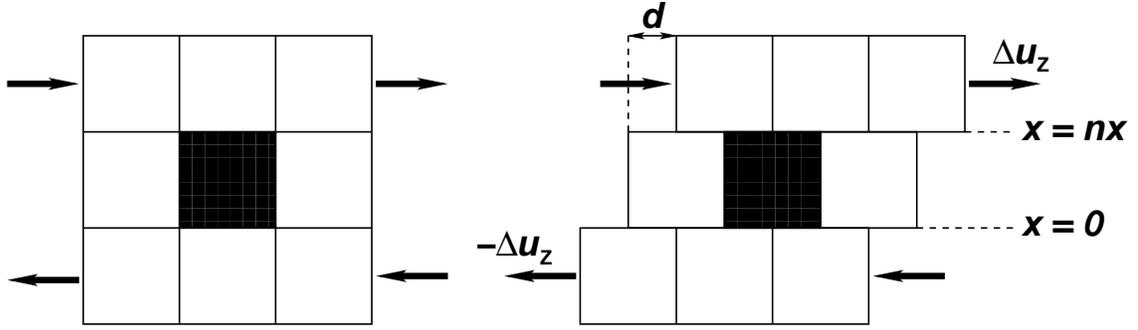}}
\caption{Our lattice-Boltzmann code LB3D implements shearing by use of
Lees-Edwards boundary conditions: particles crossing the $\bf x = 0 = nx$
boundary have their $z$-velocities altered by $\pm\Delta u_Z$ and are displaced
in the $\pm z$ direction by $d = t\Delta u_Z$ ($t$ = simulation time). The
corresponding shear rate is $s = \Delta u_Z / nx$; oscillatory shear is
achieved by setting $\Delta u_Z' = \cos(\omega t)\Delta u_Z$.  }
\label{leesedwards}
\end{figure}

The method can be described as an extension of the use of standard
periodic boundary conditions and is illustrated in
figure~\ref{leesedwards}. While with periodic boundary conditions
particles that arrive at a system boundary leave the simulation volume and
`re-enter' it on the opposite side, for a sheared system this is only true
for the boundaries not subject to shear. Particles crossing the shear
planes, which are the $\bf x = 0$ and $\bf x  = nx$ planes in our case,
get their $z$-velocities altered by $\pm\Delta u_Z$ and are displaced in
the $\pm z$ direction by $d = t\Delta u_Z$ ($t$ is the simulation time and
$nx$ is the lattice size in the $x$ direction). The corresponding shear
rate $s$ is $\Delta u_Z / nx$. This algorithm can be extended for
simulations of fluids under oscillatory shear by multiplying $\Delta u_Z$
with a time dependent cosine function of frequency $\omega$: $\Delta u_Z'
= \cos(\omega t)\Delta u_Z$.

We are applying our model to study spinodal decomposition under shear, also referred to as Couette flow. The phase
separation of binary immiscible fluids without shear has been studied in detail
by different authors and LB3D has been shown to model the underlying physics
successfully \cite{bib:gonzalez-nekovee-coveney}. 
It has been shown in the non-sheared studies of spinodal decomposition that
lattice sizes need to be large in order to overcome finite size effects, i.e.
128$^3$ has been found the minimum acceptable number of lattice sites
\cite{bib:gonzalez-nekovee-coveney}. For high shear rates, systems also have to
be very long because, if the system is too small, the domains interconnect
across the $\bf z = 0$ and $\bf z = nz$ boundary and form interconnected
lamellae in the direction of shear. Such artefacts need to be eliminated from
our simulations.

Computational steering is a very useful tool for checking on finite size
effects in an ongoing sheared fluid simulation. While being able to
constantly monitor volume-rendered colour fields or fluid densities,
the human eye turned out to be very reliable in spotting the moment when
these simulations become unphysical. In this way, we were able to keep the
computational resources required at a minimum.

From our studies we found that to avoid finite size effects 64x64x512 systems
are sufficient for low shear rates and short simulation times, but 128x128x1024
lattices are needed for higher shear rates and/or very long simulations.

The results presented in this section were all obtained for a 64x64x512
system with all relaxation times and masses set to unity, i.e.
$\tau^\sigma$=1.0, $m^\sigma$=1.0. The initial oil and water fluid
densities $f_r$ and $f_b$ were given by a random distribution between 0.0
and 0.7 (in lattice units). All simulations were performed on 64 CPUs of a
SGI Origin 3800 at CSAR in Manchester, UK.  Shear rates $s$ were set to
$0$, $7.8\times 10^{-4}$, $1.6\times 10^{-3}$, $3.1\times 10^{-3}$  and
$4.7\times 10^{-3}$ (in lattice units) in order to study the influence of
shear on the domain growth.  In order to compare between different
simulations, we define the time dependent lateral domain size $L(t)$ along
direction $i=x,y,z$ as
\begin{equation}
L_i(t)\equiv \frac{2\pi}{\sqrt{\left< k^2_i(t)\right>}},
\end{equation}
where 
\begin{equation}
\left<k^2_i(t)\right>\equiv \frac{\sum_\mathbf{k} k_i^2 S(\mathbf{k},t)}
{\sum_\mathbf{k} S(\mathbf{k},t)} 
\end{equation}
is the second order moment of the three-dimensional structure function
\begin{equation}
S(\mathbf{k},t)\equiv\frac{1}{V}\left|\phi^\prime_\mathbf{k}(t)\right|^2
\end{equation}
with respect to the cartesian component $i$. $\left< \right>$ denotes the
average in Fourier space, weighted by $S(\mathbf{k})$ and $V$ is the number of
nodes of the lattice, $\phi^\prime_\mathbf{k}(t)$ the Fourier transform of the
fluctuations of the order parameter $\phi^\prime\equiv\phi-\left<\phi\right>$,
and $k_i$ is the $i$th component of the wave vector.

\begin{figure}
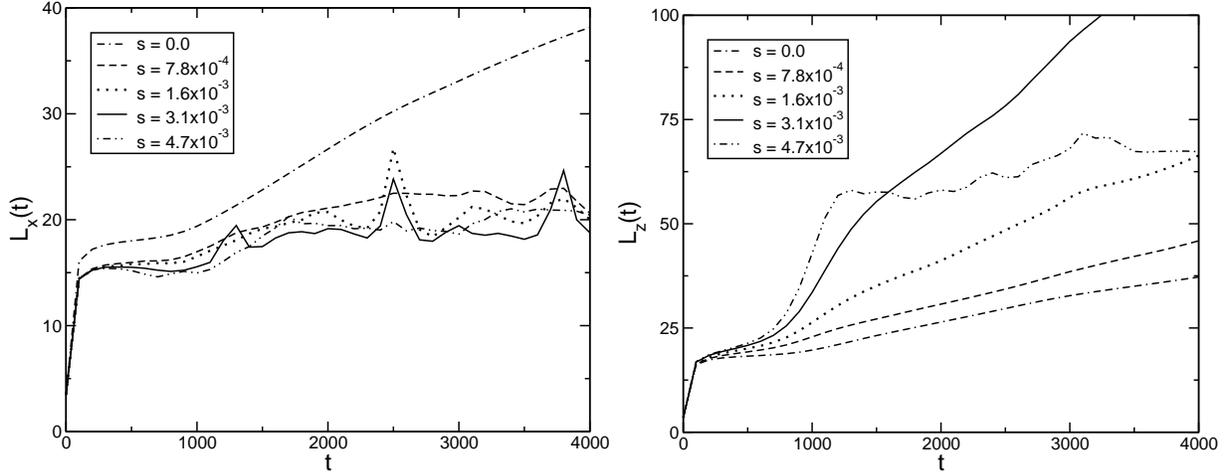

\centerline{\includegraphics[width=8cm]{Lxshearrates.eps}
\includegraphics[width=8cm]{Lzshearrates.eps}
}
\caption{The time-dependent lateral domain size $L$ in $x$ (left) and $z$
directions (right) for different shear rates. Initial fluid densities are
randomly distributed between 0 and 0.7, the system size is 64x64x512, and
$\tau^\sigma = m^\sigma =1$ (all quantities are reported in lattice units).}
\label{Lshearrates}
\end{figure}

Figure \ref{Lshearrates} shows the time dependent lateral domain size in $x$
and $z$ directions for the abovementioned parameters and up to 4000 timesteps.
At the beginning of simulations, there is a steep increase of $L$ due to
rapid diffusion of mass to nearest neighbours before the domain growth starts.
As expected, the behaviour of $L(t)$ is identical in $x$ and $z$ directions for
$s=0.0$, but is very different for $s > 0.0$.  The average slope of $L_x(t)$
decreases for increasing $s$ until phase separation almost arrests and multiple
peaks occur for $s = 1.6\times 10^{-3}$ and $s = 3.1\times 10^{-3}$ (in lattice
units). These peaks arise very regularly at approximately every 700 timesteps
in the former case and every 1500 timesteps in the latter case. For $s =
4.7\times 10^{-3}$ these peaks cannot be observed. They can be explained as
follows: if a domain reaches a substantial size, the probability of it
coalescing with a similarly sized one becomes high, but the resulting very
large domain will be unable to withstand the strain caused by the shear and
will break up a few timesteps later. For higher shear rates, domain growth in
the $x$ direction is slower than for lower shear rates and the peaks occur with
a smaller frequency. If $s$ becomes too high (as in the $s = 4.7\times 10^{-3}$
case), the imposed strain prevents domains substantially larger than the
average domain size from forming.

In the $z$ direction, shear causes elongated domains resulting in increasing
values of $L(t)$ for increasing shear rates. For $s = 4.7\times 10^{-3}$,
$L_z(t)$ grows rapidly until it saturates at $t = 1100$. A critical domain size
is reached after which domains still grow, but are very elongated and tilted by
an angle. Due to this tilting, $L_z$ saturates since it only measures the size
of the domains in $z$ direction. 

\begin{figure}
\centerline{\includegraphics[width=9cm]{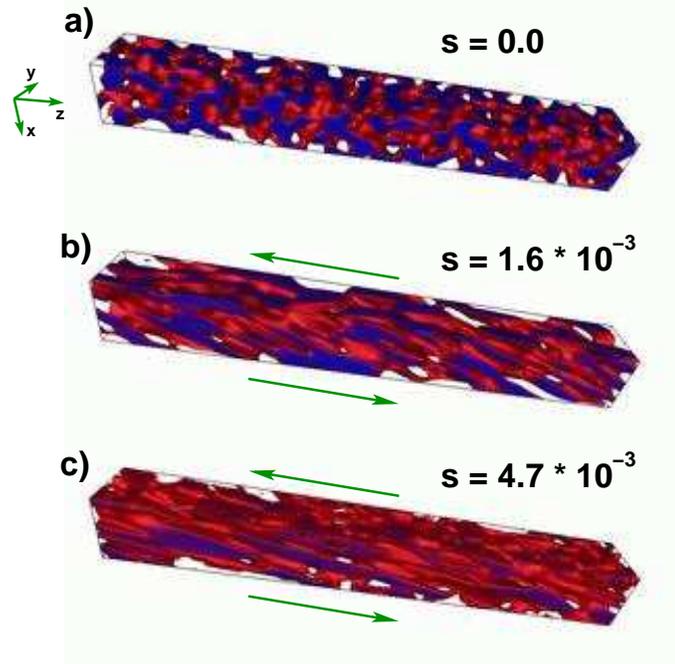}}
\caption{Volume rendered order parameter $\phi$ for increasing shear rates all
shown at timestep 3000. The system size is 64x64x512. Blue areas denote high
density of `blue' fluid and red denotes the interface between fluids. With
increasing shear rate, domains become more elongated and tilted.}
\label{shear-s0-s03}
\end{figure}

This effect is illustrated in figure \ref{shear-s0-s03} which shows
volume-rendered snapshots of the order parameter $\phi$ for shear rates $s =
0.0$, $1.6\times 10^{-3}$, and $4.7\times 10^{-3}$, all at timestep 3000.
Areas of high density of `blue' fluid are coloured blue and the interface
between both fluids is coloured red. The figure shows how the domains develop
evenly in $x$ and $z$ direction for $s = 0.0$, but become tilted and elongated
under shear. This effect increases with increasing shear rate, resulting in
very long and slim lamellae in the $s = 4.7\cdot 10^{-3}$ case. 

The results presented in this section are preliminary. We hope to be able
to report on more detailed studies of essentially finite-size-free
simulations of complex fluid mixtures under shear in the near future. We
are planning to utilize our code to quantify domain growth and compare our
results to previous theoretical work and experimental results
\cite{bib:cates-kendon-bladon-desplat,bib:wagner-yeomans-shear,bib:corberi-gonnella-lamura}.
Applying oscillatory instead of constant shear is a natural extension of
this work and simulations are already ongoing
\cite{bib:xu-gonnella-lamura}. We are also studying the
properties of amphiphilic fluid mixtures under shear including the effect
shear has on previously formed cubic mesophases such as the 'P'-phase
\cite{bib:nekovee-coveney} and the gyroid cubic mesophase
\cite{bib:gonzalez-coveney}. Such complex fluids are expected to exhibit
non-Newtonian properties.

\section{Multiphase flow in porous media}
\label{Sec:Porous}
The study of transport phenomena in porous media is of great interest in
fields ranging from oil recovery and water purification to industrial
processes like catalysis.  In particular, the oilfield industry uses
complex, non-Newtonian, multicomponent fluids (containing polymers,
surfactants and/or colloids, brine, oil and/or gas), for processes like
fracturing, well stimulation and enhanced oil recovery.  The rheology and
flow behaviour of these complex fluids in a rock is different from their
bulk properties.  It is therefore of considerable interest to be able to
characterise and predict the flow of these fluids in porous media.

The flow of a single phase non-Newtonian fluid through two-dimensional
porous media has been addresses with lattice-Boltzmann methods
\cite{bib:aharonov-rothman,bib:chin-boek-coveney}, using a `top-down'
approach, in which the effective dynamic viscosity of the fluid, and hence
the relaxation parameter in the BGK equation \ref{eq:lbgk}, explicitly
depends on the strain rate tensor through a power law.

However, from the point of view of a modelling approach, the treatment of
complex fluids in three-dimensional complex geometries is an ambitious
goal.  In this paper we shall only consider binary (oil/water) mixtures of
Newtonian fluids, since this is a first and necessary step in the
understanding of multiphase fluid flow in porous media.

The advantage of using lattice-Boltzmann (or lattice-gas) techniques in
studying flow in porous media is that complex geometries can be easily
implemented and the flow problem solved therein, since the evolution of
the particle distribution functions can be described in terms of local
collisions with the obstacle sites using simple bounce-back boundary
conditions.  Synchrotron based X-ray microtomography (XMT) imaging
techniques provide high resolution, three-dimensional digitised images of
rock samples.  By using the lattice-Boltzmann approach in combination with
these high resolution images of rocks, not only it is possible to compute
macroscopic transport coefficients, such as the permeability of the
medium, but information on local fields, such as velocity or fluid
densities, can also be obtained at the pore scale thus providing a
detailed insight into local flow characterisation and supporting the
interpretation of experimental measurements \cite{bib:auzeraisGRL96}.

The XMT technique measures the linear attenuation coefficient from which
the mineral concentration and composition of the rock can be computed.
From the tomographic image of the rock volume the topology of the void
space can be derived, such as pore size distribution and tortuosity, and
the permeability and conductivity of the rock can be computed
\cite{bib:spannePRL96}.  The tomographic data are represented by a
reflectivity greyscale value and are arranged in voxels in a three
dimensional coordinate system.  The linear size of each voxel is defined
by the imaging resolution, which is usually on the order of microns.  By
introducing a threshold to discriminate between pore sites and rock sites,
these greyscale images can be reduced to a binary (0's and 1's)
representation of the rock geometry.  Using the lattice-Boltzmann method,
single phase or multiphase flow can then be described in these real porous
media.

Lattice Boltzmann and lattice gas techniques have already been applied to
study single and multiphase flow through three-dimensional
microtomographic reconstruction of porous media.  For example, Martys and
Chen \cite{bib:martys-chen} and Ferr{\'e}ol and Rothman
\cite{bib:ferreol-rothman} studied relative permeabilities of binary
mixtures in Fontainebleau sandstone.  These studies validated the model
and the simulation techniques, but were limited to small lattice sizes, of
the order of $64^3$.

The possibility of describing fluid flow in real rock samples gives the
advantage of being able to make comparisons with experimental results
obtained on the same, or similar, pieces of rock.  Of course, to achieve a
reasonable comparison, the size of the rock used in lattice-Boltzmann
simulations should be of the same order of magnitude as the system used in
the experiments, or at least large enough to capture the rock's
topological features.  The more inhomogeneous the rock, the larger the
sample size needs to be in order to describe the correct pore distribution
and connectivity.  Another reason for needing to use large lattice sizes
is the influence of boundary conditions and lattice resolution on the
accuracy of the lattice-Boltzmann method.  It has been shown (see for
example \cite{bib:He97}, \cite{bib:chen-doolen} and references therein) that if
the Bhatnagar-Gross-Krook (BGK) \cite{bib:bgk} approximation of the
lattice-Boltzmann equation is used, the so-called bounce-back boundary
condition at the wall sites actually mimics boundaries which move with a
speed that depends on the relaxation parameter $\tau$ of the collision
operator in the BGK equation (\ref{eq:lbgk}).  The relaxation parameter
determines the kinematic viscosity of the simulated fluid.  This implies
that the computed permeability is a function of the viscosity.

The accuracy of lattice-Boltzmann simulations also depends on the Knudsen
number which represents the ratio of the mean free path of the fluid
particles and the characteristic length scale of the system (such as the
pore diameter). To accurately describe hydrodynamic behaviour this ratio
has to be small. If the pores are resolved with an insufficient number of
lattice points finite size effects arise, leading to an inaccurate
description of the flow field.

The error in solving the flow field increases with increasing viscosity
(relaxation time), but this viscosity dependence becomes weak with
increasing lattice resolution. Hence it is desirable to use a high
resolution within the pore space in order to decrease the error induced by
the use of bounce-back boundary conditions. However, increasing the
resolution means increasing the lattice size, hence the computational cost
of the simulation. 

Boundary conditions other than bounce-back have been proposed and shown to
give correct velocities at the boundaries.  However these methods are
either suitable only for flat interfaces \cite{bib:inamuro} or they are
cumbersome to implement \cite{bib:verberg-ladd-01}, reducing the
efficiency of the lattice-Boltzmann method.

Large lattices require a highly scalable code, access to high performance
computing, terascale storage facilities and high performance
visualisation.  LB3D provides the first of these, while the others are now
offered by the UK High Performance Computing services, and are also
accessible via the UK e-Science Grid using RealityGrid capabilities.

\begin{figure}
\centerline{\includegraphics[width=0.5\textwidth]{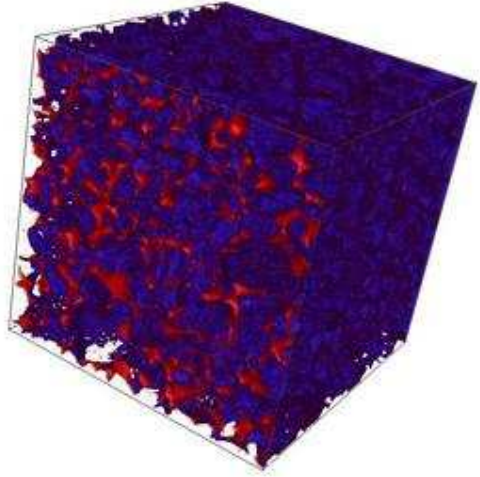}}
\caption{
Rendering of an X-ray microtomographic image of a 
$512^3$ sample of Bentheimer sandstone.
The data have a resolution of $4.9\mu m$.
The pore space is shown in red, while the rock is represented in blue.} 
\label{f:Benth}
\end{figure}

Using LB3D and capability computing services provided by Manchester CSAR
SGI Origin 3800, we were able to simulate drainage and imbibition
processes in a $512^3$ subsample of Bentheimer sandstone X-ray tomographic
data. The whole set of XMT data represented the image of a Bentheimer
sandstone of cylindrical shape with diameter 4mm and length 3mm.  The
XMT data were obtained at the European Synchrotron Research Facility
(Grenoble) at a resolution of $4.9 \mu {\rm m}$, resulting in a data set
of approximately 816x816x612 voxels. Figure \ref{f:Benth} shows a snapshot of the
$512^3$ subsystem.

The aim of this study is to compare velocity distributions with the ones
measured by magnetic resonance imaging (MRI) of  oil and brine
infiltration into saturated Bentheimer rock core \cite{MRISheppard}.  The
rock sample used in these MRI experiments had a diameter of 38 mm and was
70 mm long; three-dimensional images of the rock were acquired at a
resolution of 280 microns. The system size we used in lattice-Boltzmann
simulations was smaller than the sample used in MRI experiments, but still
of a similar order of magnitude and large enough to represent the rock
geometry. On the other hand, the higher space resolution provided by the
lattice-Boltzmann method permits a detailed characterisation of the flow
field in the pore space, hence providing a useful tool to interpret the
MRI experiments, for example in identifying regions of stagnant fluid.
Detailed results of these lattice-Boltzmann simulations will be presented
in a future paper.

\subsection{Binary flow in a digitised Bentheimer rock sample.}
Darcy's law \cite{bib:darcy} for binary immiscible (oil/water) fluid mixtures
was investigated and the dependence of the relative permeability
coefficients on water saturation was derived and compared with lattice-gas
studies.  The extended Darcy's law for binary flow takes the form 
\begin{equation}
{\bf J}_i=\sum_{j=1}^2 k_{ij}(S)\frac{k}{\mu_i}{\bf X}_j, \hspace{0.8cm} (i=1,2)
\end{equation}
where ${\bf J}_i$ is the flux of the $i$th component and ${\bf X}_j$ is
the force acting on the $j$th component. $k_{ij}(S)$ is the relative
permeability coefficient depending on the saturation $S$, $k$ is the
permeability of the medium and $\mu_i$ the viscosity of component $i$.

Using the lattice-Boltzmann method it is easy to selectively force only
one component in a binary mixture, leaving the other one unforced.  In
this way the diagonal terms of the relative permeability matrix ($k_{ii}$)
can be computed by analysing the flux of one component when it is forced,
while the cross terms ($k_{ij}$) can be computed from the flux of one
component when the other one is forced.

\begin{figure}
\centerline{\includegraphics[width=0.5\textwidth]{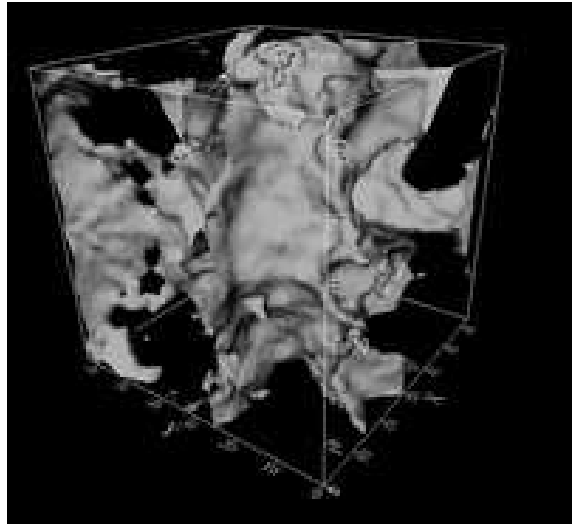}}
\caption{
A $64^3$ sample of Bentheimer sandstone at a lattice resolution of $4.9\mu m$.
The figure shows the isosurface delimiting the pore volume. 
The pore space is in the inner region of the isosurface.
}
\label{f:64}
\end{figure}

Since we want to study the flow behaviour for different forcing levels and
for forcing applied in turn to both fluids, a large number of simulations
is required.  Hence we limited the size of our system to a subsample of
64x64x32 voxels, mirrored in the $z$-direction (flow direction) to give a
final size of $64^3$ lattice sites (see figure \ref{f:64}). Periodic
boundary conditions were applied in all directions.
Immiscible, binary mixtures of oil and water were flowed in this sample,
at different forcing levels and by forcing in turn either the water or the
oil component.  In each of these numerical experiments, the system was
initialised with a 50:50 mixture of oil and water, both given the same
viscosity and the same initial density distributions.  The rock walls were
made fully water wettable, to reproduce experimental conditions and to
discriminate between the two fluids which would otherwise be equivalent.
The rock wettability is implemented by assigning each rock site a density
distribution equal to the initial density of water.  These density
distributions do not flow, but exert a (repulsive) force on the oil
component, pushing it away from the rock walls.

As the two components flow, they initially phase separate and, after some
time, a steady state is reached.  Here we are only interested in the
velocity field at the steady state. If the total time allocated for the
simulation is longer than the time needed to reach the steady state, we
would waste CPU time performing calculations which are useless.  If we run
a simulation for less time than needed to reach the steady state, using
the checkpoint-restart facility within LB3D we can resume the simulation
and continue the run until the steady state is reached.  In both cases,
steering  improves the efficiency of the runs.  By steering we can dump
velocity and density fields at variable frequency, check whether the
fields at two different times differ or not, and in case their difference
is less than a given threshold, decide that the steady state is reached
and stop the simulation.  In the simulations presented here an average of
15000 time steps is needed to reach the steady state.  At this time the
total flux (normalised by the pore volume) can be computed.

Figure \ref{f:Kii_Kij:a} shows the force/flux dependence for the forced
fluids.  The linearity of force/flux holds for all the forcing levels
considered.  At any given forcing level, the wetting fluid flows less than
the non-wetting one.  This is due to the fact that the wetting fluid
interacts with the rock walls, and adheres to them, hence exerting more
resistance to flow, while the non-wetting fluid is lubricated by the
wetting fluid.  Similar results have been achieved in three-dimensional
lattice-gas studies of binary flow in Fontainebleau sandstone
\cite{bib:love-maillet-coveney,bib:olson-rothman}.  A difference between
our results and the lattice-gas ones is that in the latter the authors
observed the presence of a capillary threshold, a minimum forcing level
required to make the non-wetting fluid flow, while here we observe flow
even at small forcing levels.  The presence of this threshold is due to
geometric constraints imposed by the distribution and size of rock pores
and throats, which can trap bubbles of the non wetting fluid.  In the rock
sample we used for this study there are no such narrow throats, hence we
would not expect to observe any capillary threshold.

In figure \ref{f:Kii_Kij:b} the force/flux relation is plotted for the
non-forced fluids.  In this case at any applied forcing the non wetting
fluid flows more than the wetting one.  For the non wetting fluid we
observe viscous coupling, i.e.  the fluid flows even if it is not directly
forced, and Darcy's linearity is found for all forcing levels.  On the
other hand, the wetting fluid does not flow until a sufficiently high
force is applied. This is due to the capillary forces.  Hence Darcy's law
is observed to hold only for sufficiently high forcing.

From the linear regime regions in both graphs we computed the relative
permeability coefficients: $k_{\rm ww}=1.6$, $k_{\rm oo}=3.3$, $k_{\rm
ow}=0.77 $, $k_{\rm wo}=0.57$, where the subscripts indicate water (w) and
oil (o).  The diagonal terms are one order of magnitude larger than the
cross terms, which can be expected because the cross terms represent the
response of one fluid when the {\it other one} is forced.  This is also in
agreement with the results from lattice-gas studies
\cite{bib:love-maillet-coveney}.

A previous much debated issue is whether the off-diagonal terms in the
relative permeability matrix should satisfy a reciprocity relationship.
The reciprocity of the coefficients in macroscopic linear transport laws
of the form
\begin{equation}
J_i = \sum_j L_{ij}X_j,
\end{equation}
where $J_i$ is a current and $X_j$ is a force conjugate to the current, is
a consequence of Onsager's  regression hypothesis and it holds for systems
which are linearly perturbed from equilibrium \cite{bib:groot-mazur}. 

Our results show a linear dependence between force and flow, but the
off-diagonal coefficients we obtained from the linear regime region have
slightly different values for the wetting and non-wetting fluids.  For
binary immiscible fluid flow in porous media, where complex interfacial
dynamics plays a major role, it is not clear if Onsager's theory can be
applied.  In the aforementioned  lattice-gas studies, Onsager's
reciprocity was found, but no clear theoretical justification has been
given for the reason that this should hold under the general nonlinear
conditions pertaining {\cite{bib:flekkoy-pride}.

It can also be observed that there are no error bars provided with our
results.  This is due to the fact that in lattice-Boltzmann, as opposite
to lattice-gas simulations, there is no noise, and indeed this is one of
the major computational advantages of the method.  Nevertheless, we plan
to perform the same set of simulations starting from different initial
conditions, which may lead to different fluid/fluid interfacial structures
and fluid transport coefficients, and then derive error bars from the flow
computed in each of these simulations.  More studies on different pore
space geometries and larger systems need to be done to address the general
validity of Darcy's law for binary mixtures and Onsager's reciprocity
hypothesis. 

\begin{figure}
\subfigure[]{\includegraphics[width=0.5\textwidth]{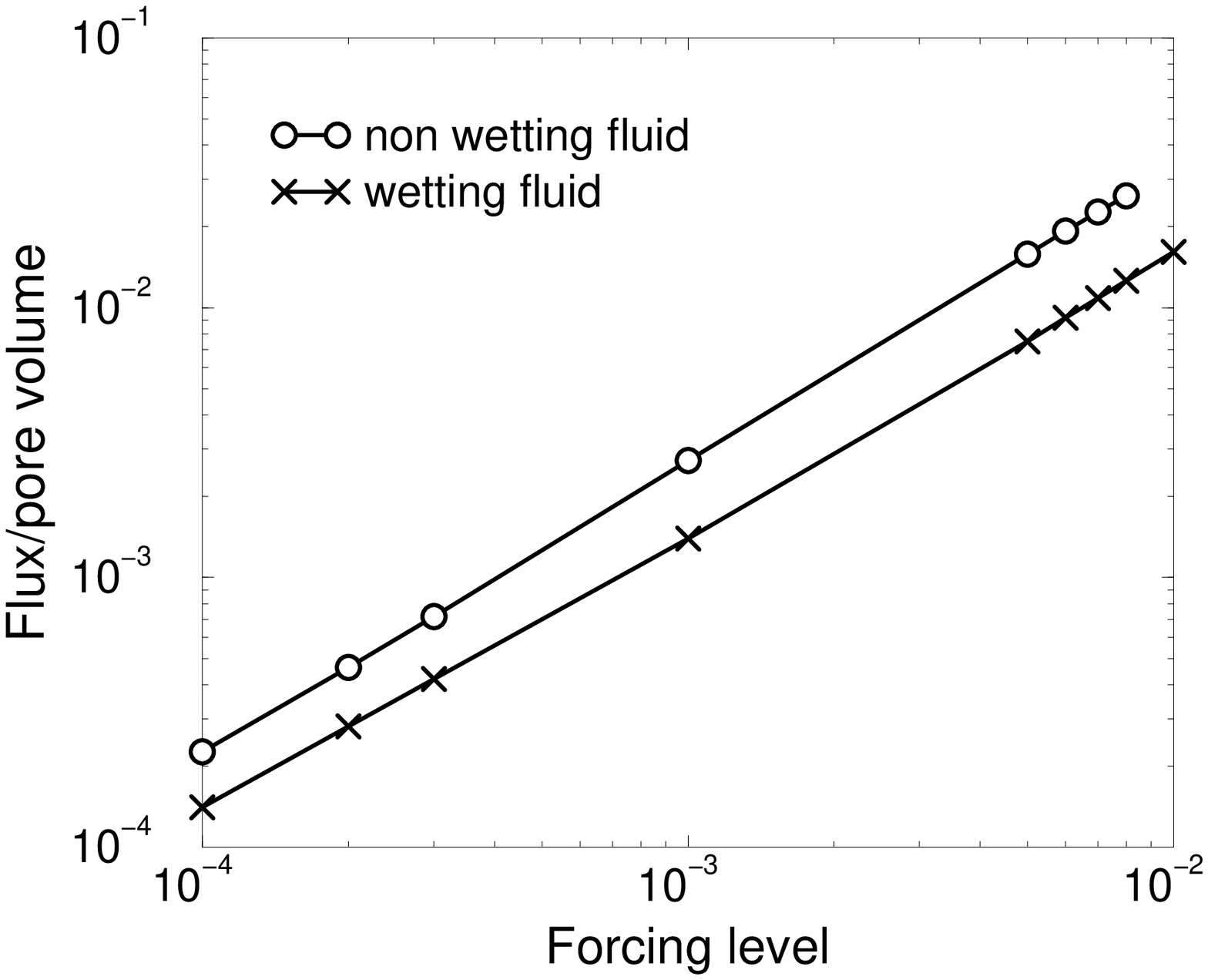}\label{f:Kii_Kij:a}}
\subfigure[]{\includegraphics[width=0.5\textwidth]{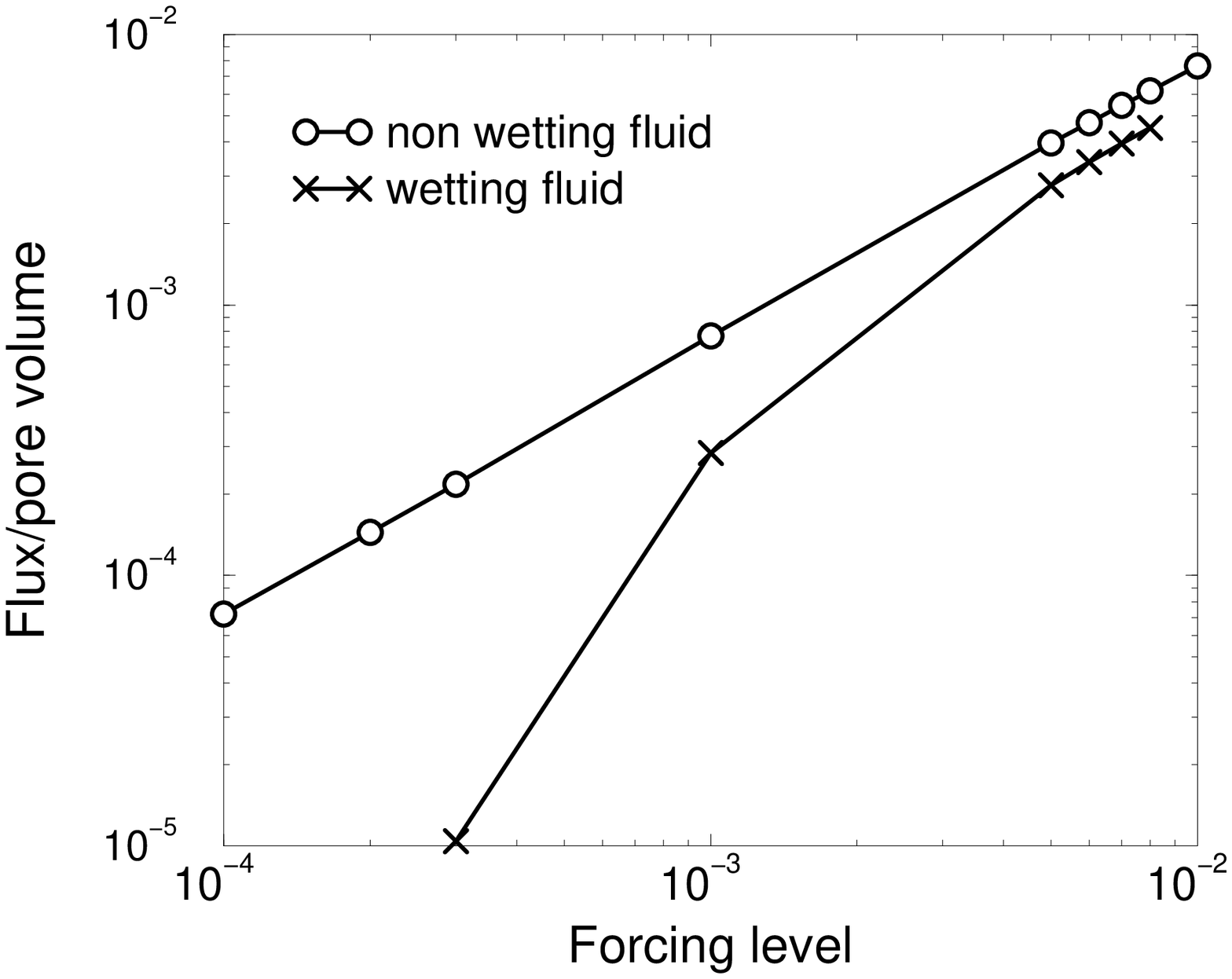}\label{f:Kii_Kij:b}}
\caption{
Flux of binary fluid when forced (a) and when unforced (b).
All quantities are given in lattice units and the flux is normalised by the pore volume.
}
\label{f:Kii_Kij}
\end{figure}

\section{Conclusions}
\label{Sec:Conclusion}
This paper describes optimal implementation of large scale
lattice-Boltzmann simulations of two-phase fluid dynamics through
exploitation of capability computing and computational steering.
Capability computing promotes the execution of scalable codes that utilise
a large fraction and sometime the entire allocation of processors on a big
supercomputer.  Computational steering then ensures that this massive set
of resources is used optimally to generate meaningful scientific data.  We
illustrated the use of these approaches by reporting preliminary results
from two applications which benefit substantially from large scale
simulation.  The first of these was concerned with Couette (shear) flow,
where simulation cells of high aspect ratio are needed to eliminate finite
size effects; the second described two-phase flow in large portions of
digitised data obtained from X-ray microtomographic studies of Bentheimer
sandstone.  The most efficient utilisation of computational steering of
such large scale simulations utilises a computational grid.  These grids
are in their infancy today and much more work needs to be done to render
them transparent to users.  Nevertheless, important advances have already
been made, and here we described the grid-enablement of our
lattice-Boltzmann codes.  Our experience with grids to date leads us to
conclude that much lighter middleware solutions will be required to foster
their widespread use.
\section*{Acknowledgements}
We would like to thank J.~Chin, N.~Gonz\'{a}lez-Segredo, and S.~Jha
(University College London), A.R.~Porter, and S.M.~Pickles (University of
Manchester), E.S.~Boek, and J.~Crawshaw (Schlumberger Cambridge Research)
for fruitful discussions and E.~Breitmoser from the Edinburgh Parallel
Computing Centre for her contribution to the Lees-Edwards routines in our
code. We are grateful to EPSRC for funding much of this research through
RealityGrid grant GR/R67699 and for providing access to SGI Origin 3800,
and Origin 2000 supercomputers at Computer Services for Academic Research
(CSAR), Manchester, UK.

We acknowledge the European Synchrotron Radiation Facility for provision
of synchrotron radiation facilities and we would like to thank Dr Peter
Cloetens for assistance in using beamline ID19, as well as Professor Jim
Elliott and Dr Graham Davis of Queen Mary, University of London, for their
work in collecting the raw data and reconstructing the x-ray
microtomography data sets used in our Bentheimer sandstone images.


\begin{thebibliography}{}

\bibitem[\protect\astroncite{Aharonov and Rothman}{1993}]{bib:aharonov-rothman}
Aharonov, E. and Rothman, D.~H.: 1993,
\newblock {\em Geophys. Research Lett.} {\bf 20}, 679

\bibitem[\protect\astroncite{Auzerais {\it et~al.}}{1996}]{bib:auzeraisGRL96}
Auzerais, F.~M., Dunsmuir, J., Ferreol, B.~B., Martys, N., Olson, J.,
  Ramakrishnan, T., Rothman, D., and Schwartz, L.: 1996,
\newblock {\em Geophys. Research Lett.} {\bf 23(7)}, 705

\bibitem[\protect\astroncite{Berman {\it et~al.}}{2003}]{bib:Berman}
Berman, F., Fox, G., and Hey, T.: 2003,
\newblock {\em Grid Computing: Making The Global Infrastructure a Reality},
\newblock John Wiley and Sons

\bibitem[\protect\astroncite{Bhatnagar {\it et~al.}}{1954}]{bib:bgk}
Bhatnagar, P.~L., Gross, E.~P., and Krook, M.: 1954,
\newblock {\em Phys. Rev.} {\bf 94(3)}, 511

\bibitem[\protect\astroncite{Brooke {\it
  et~al.}}{2003}]{bib:brooke-coveney-harting}
Brooke, J.~M., Coveney, P.~V., Harting, J., Jha, S., Pickles, S.~M., Pinning,
  R.~L., and Porter, A.~R.: 2003,
\newblock in {\em Proceedings of the UK e-Science All Hands Meeting, September
  2-4}, pp 885--889,
\newblock (http://www.nesc.ac.uk/events/ahm2003/AHMCD/pdf/179.pdf)

\bibitem[\protect\astroncite{Cates {\it
  et~al.}}{1999}]{bib:cates-kendon-bladon-desplat}
Cates, M.~E., Kendon, V.~M., Bladon, P., and Desplat, J.~C.: 1999,
\newblock {\em Faraday Disc.} {\bf 112}, 1

\bibitem[\protect\astroncite{Chapman and Cowling}{1952}]{bib:chapman-cowling}
Chapman, S. and Cowling, T.~G.: 1952,
\newblock {\em The Mathematical Theory of Non-Uniform Gases},
\newblock Cambridge University Press, second edition

\bibitem[\protect\astroncite{Chen {\it
  et~al.}}{2000}]{bib:chen-boghosian-coveney}
Chen, H., Boghosian, B.~M., Coveney, P.~V., and Nekovee, M.: 2000,
\newblock {\em Proc. R. Soc. Lond. A} {\bf 456}, 2043

\bibitem[\protect\astroncite{Chen and Doolen}{1998}]{bib:chen-doolen}
Chen, S. and Doolen, G.~D.: 1998,
\newblock {\em Annu. Rev. Fluid Mech.} {\bf 30}, 329

\bibitem[\protect\astroncite{Chin {\it et~al.}}{2002}]{bib:chin-boek-coveney}
Chin, J., Boek, E.~S., and Coveney, P.~V.: 2002,
\newblock {\em Proc. R. Soc. Lond. A} {\bf 360}, 547

\bibitem[\protect\astroncite{Chin {\it et~al.}}{2003}]{bib:chin-harting-jha}
Chin, J., Harting, J., Jha, S., Coveney, P.~V., Porter, A.~R., and Pickles,
  S.~M.: 2003,
\newblock {\em Contemporary Physics} {\bf 44(5)}, 417

\bibitem[\protect\astroncite{Corberi {\it
  et~al.}}{2002}]{bib:corberi-gonnella-lamura}
Corberi, F., Gonnella, G., and Lamura, A.: 2002,
\newblock {\em Phys. Rev. E} 66(016114)

\bibitem[\protect\astroncite{Coveney}{2003}]{bib:coveney-nobel}
Coveney, P.~V.: 2003,
\newblock {\em Phil. Trans. R. Soc. Lond. A} {\bf 361(1807)}, 1057

\bibitem[\protect\astroncite{Darcy}{1856}]{bib:darcy}
Darcy, H.: 1856,
\newblock {\em Les Fontaines Publiques de la Ville de Dijon},
\newblock Dalmont, Paris

\bibitem[\protect\astroncite{de~Groot and Mazur}{1985}]{bib:groot-mazur}
de~Groot, S.~R. and Mazur, P.: 1985,
\newblock {\em Nonequilibrium thermodynamics},
\newblock Dover Publications Inc., New York

\bibitem[\protect\astroncite{Ferr{\'e}ol and
  Rothman}{1995}]{bib:ferreol-rothman}
Ferr{\'e}ol, B. and Rothman, D.~H.: 1995,
\newblock {\em Transport in porous media} {\bf 20}, 3

\bibitem[\protect\astroncite{Flekk{\o}y and Pride}{1999}]{bib:flekkoy-pride}
Flekk{\o}y, E.~G. and Pride, S.~E.: 1999,
\newblock {\em Phys. Rev. E} {\bf 60(4)}, 4130

\bibitem[\protect\astroncite{Foster and Kesselman}{1999a}]{gridbook2}
Foster, I. and Kesselman, C.: 1999a,
\newblock in I. Foster and C. Kesselman (eds.), {\em The Grid: Blueprint for a
  New Computing Infrastructure}, pp 15--25, Morgan Kaufmann

\bibitem[\protect\astroncite{Foster and Kesselman}{1999b}]{globus}
Foster, I. and Kesselman, C.: 1999b,
\newblock in I. Foster and C. Kesselman (eds.), {\em The Grid: Blueprint for a
  New Computing Infrastructure}, p. 259, Morgan Kaufmann

\bibitem[\protect\astroncite{Gonz\'{a}lez-Segredo and
  Coveney}{2003}]{bib:gonzalez-coveney}
Gonz\'{a}lez-Segredo, N. and Coveney, P.~V.: 2003,
\newblock {\em e-print: www.arXiv.org/cond-mat/0310390}

\bibitem[\protect\astroncite{Gonz\'{a}lez-Segredo {\it
  et~al.}}{2003}]{bib:gonzalez-nekovee-coveney}
Gonz\'{a}lez-Segredo, N., Nekovee, M., and Coveney, P.~V.: 2003,
\newblock {\em Phys. Rev. E} 67(046304)

\bibitem[\protect\astroncite{Harting {\it
  et~al.}}{2003}]{bib:harting-wan-coveney}
Harting, J., Wan, S., and Coveney, P.~V.: 2003,
\newblock {\em Capability Computing: The newsletter of the HPCx community} 2

\bibitem[\protect\astroncite{He {\it et~al.}}{1997}]{bib:He97}
He, X., Q.~Zou, L.~L., and Dembo, M.: 1997,
\newblock {\em J. Stat. Phys.} {\bf 87(115)},

\bibitem[\protect\astroncite{Inamuro {\it et~al.}}{1995}]{bib:inamuro}
Inamuro, T., Yoshino, M., and Ogino, F.: 1995,
\newblock {\em Phys. Fluids} {\bf 7(12)}, 2928

\bibitem[\protect\astroncite{Kendon {\it
  et~al.}}{2001}]{bib:kendon-cates-pagonabarraga-desplat-bladon}
Kendon, V.~M., Cates, M.~E., Pagonabarraga, I., Desplat, J.~C., and Bladon, P.:
  2001,
\newblock {\em J. Fluid Mech.} {\bf 440}, 147

\bibitem[\protect\astroncite{Lees and Edwards}{1972}]{bib:lees-edwards}
Lees, A.~W. and Edwards, S.~F.: 1972,
\newblock {\em J. Phys. C.} {\bf 5(15)}, 1921

\bibitem[\protect\astroncite{Liboff}{1990}]{bib:liboff}
Liboff, R.~L.: 1990,
\newblock {\em Kinetic Theory: Classical, Quantum, and Relativistic
  Descriptions},
\newblock Prentice-Hall

\bibitem[\protect\astroncite{Love {\it
  et~al.}}{2001}]{bib:love-maillet-coveney}
Love, P.~J., Maillet, J., and Coveney, P.~V.: 2001,
\newblock {\em Phys. Rev. E} 64(061302)

\bibitem[\protect\astroncite{Love {\it
  et~al.}}{2003}]{bib:love-nekovee-coveney-chin-gonzalez-martin}
Love, P.~J., Nekovee, M., Coveney, P.~V., Chin, J., Gonz\'{a}lez-Segredo, N.,
  and Martin, J. M.~R.: 2003,
\newblock {\em Comp. Phys. Comm.} {\bf 153(3)}, 340

\bibitem[\protect\astroncite{Martys and Chen}{1996}]{bib:martys-chen}
Martys, N.~S. and Chen, H.: 1996,
\newblock {\em Phys. Rev. E} {\bf 53(1)}, 743

\bibitem[\protect\astroncite{Nekovee {\it
  et~al.}}{2001}]{bib:nekovee-chin-gonzalez-coveney}
Nekovee, M., Chin, J., Gonz\'{a}lez-Segredo, N., and Coveney, P.~V.: 2001,
\newblock in {E. Ramos {\em et al}} (ed.), {\em Computational Fluid Dynamics,
  Proceedings of the Fourth UNAM Supercomputing Conference, Singapore}, pp
  204--212, World Scientific

\bibitem[\protect\astroncite{Nekovee and Coveney}{2001}]{bib:nekovee-coveney}
Nekovee, M. and Coveney, P.~V.: 2001,
\newblock {\em J.~Am.~Chem.~Soc.} {\bf 123(49)}, 12380

\bibitem[\protect\astroncite{Olson and Rothman}{1997}]{bib:olson-rothman}
Olson, J.~F. and Rothman, D.~H.: 1997,
\newblock {\em J. Fluid Mech.} {\bf 341}, 343

\bibitem[\protect\astroncite{Shan and Chen}{1993}]{bib:shan-chen}
Shan, X. and Chen, H.: 1993,
\newblock {\em Phys. Rev. E} {\bf 47(3)}, 1815

\bibitem[\protect\astroncite{Sheppard {\it et~al.}}{2003}]{MRISheppard}
Sheppard, S., Mantle, M., Sederman, A., Johns, M., and Gladden, L.~F.: 2003,
\newblock {\em Magnetic Resonance Imaging} {\bf 21}, 365

\bibitem[\protect\astroncite{Spanne {\it et~al.}}{1994}]{bib:spannePRL96}
Spanne, P., Thovert, J.~F., Jacquin, C.~J., Lindquist, W.~B., Jones, K.~W., and
  Adler, P.~M.: 1994,
\newblock {\em Phys. Rev. Lett.} {\bf 73(14)}, 2001

\bibitem[\protect\astroncite{Succi}{2001}]{bib:succi}
Succi, S.: 2001,
\newblock {\em The Lattice {B}oltzmann Equation for Fluid Dynamics and Beyond},
\newblock Oxford University Press

\bibitem[\protect\astroncite{Verberg and Ladd}{2001}]{bib:verberg-ladd-01}
Verberg, R. and Ladd, A. J.~C.: 2001,
\newblock {\em Phys. Rev. E} {\bf 65},

\bibitem[\protect\astroncite{Wagner and
  Pagonabarraga}{2002}]{bib:wagner-pagonabarraga}
Wagner, A.~J. and Pagonabarraga, I.: 2002,
\newblock {\em J. Stat. Phys.} {\bf 107}, 521

\bibitem[\protect\astroncite{Wagner and
  Yeomans}{1999}]{bib:wagner-yeomans-shear}
Wagner, A.~J. and Yeomans, J.~M.: 1999,
\newblock {\em Phys. Rev. E} {\bf 59(4)}, 4366

\bibitem[\protect\astroncite{Xu {\it et~al.}}{2003}]{bib:xu-gonnella-lamura}
Xu, A., Gonnella, G., and Lamura, A.: 2003,
\newblock {\em Phys. Rev. E} 67(056105)

\end{thebibliography}

\end{document}